\documentclass[twoside,a4paper]{article}
\usepackage{amsmath,graphicx,amssymb,fancyhdr,amsthm,textcomp,hyperref}
\usepackage{float}
\usepackage{subcaption}
\usepackage[english]{babel}
\usepackage[numbers]{natbib}
\usepackage[acronym, nopostdot,nomain,nonumberlist,numberedsection]{glossaries}

\usepackage[utf8]{inputenc}

\usepackage{amsmath,graphicx,amssymb,fancyhdr,amsthm,textcomp} 
\newtheorem{thm}{Theorem}[section]

\theoremstyle{definition} 
  
\theoremstyle{remark}  
\newtheorem{rem}[thm]{Remark}  
\def\beq{\begin{eqnarray}}  
\def\eeq{\end{eqnarray}}  
\def\bsp{\begin{split}}  
\def\esp{\end{split}}

\newcommand{\mbold}[1]{\mbox{\boldmath{\ensuremath{#1}}}}

 \def \bell {\mbox{{\mbold\ell}}}
\def \bn {\mbox{{\bf n}}}

\def\Ep{\mathcal{E}}
 
\begin{document}  
  

\title{\Large\textbf{An invariant characterization of the quasi-spherical Szekeres dust models}}  
\author{{\large\textbf{A. A. Coley$^{\heartsuit}$, N. Layden$^{\heartsuit}$ and D. D. McNutt$^{\clubsuit}$  }}
\vspace{0.3cm} \\ 
$^{\heartsuit}$Department of Mathematics and Statistics,\\
Dalhousie University,
Halifax, Nova Scotia,\\
Canada B3H 3J5\\
$^{\clubsuit}$ Faculty of Science and Technology,\\
University of Stavanger, 
N-4036 Stavanger, Norway  \\
\vspace{0.3cm} \\
\texttt{}}
\date{}  
\maketitle  
\pagestyle{fancy}  
\fancyhead{} 
\fancyhead[EC]{}  
\fancyhead[EL,OR]{\thepage}  
\fancyhead[OC]{}  
\fancyfoot{} 
  
  \begin{abstract}

The quasi-spherical Szekeres dust solutions are a generalization of the spherically symmetric Lemaitre-Tolman-Bondi dust models where the spherical shells of constant mass are non-concentric. The quasi-spherical Szekeres dust solutions can be considered as cosmological models and are potentially models for the formation of primordial black holes in the early universe. Any collapsing quasi-spherical Szekeres dust solution where an apparent horizon covers all shell-crossings that will occur can be  considered as a model for the formation of a black hole. In this paper we will show that the apparent horizon can be detected by a Cartan invariant. We will show that particular Cartan invariants characterize properties of these solutions which have a physical interpretation such as: the expansion or contraction of spacetime itself, the relative movement of matter shells, shell-crossings and the appearance of necks and bellies.  

  \end{abstract}

\section{Introduction}

The Szekeres solutions belong to a larger class of solutions known in the literature as {\it silent universes}, due to the matter source being a perfect fluid without pressure, (i.e., dust) and the vanishing of the magnetic Weyl tensor. The latter condition implies that there cannot be gravitational waves propagating through space \cite{zak}. Each point in a silent universe evolves on its own without being affected by other regions. This can be seen more explicitly by noting that the non-linear partial differential equations of general relativity (GR) can be decoupled into a system of ordinary differential equations dictating the evolution of the physical quantities describing the system (i.e., such as the expansion rate, the shear tensor, the electric Weyl tensor and the energy density) \cite{Wainwright}; the lack of spatial derivatives in these equations ensure that different regions of space will not affect each other and simplifies the analysis of the silent universes with regards to structure formation in inhomogeneous and anisotropic cosmological models \cite{bolejko}. 

In cosmology it is believed that structure formation arises from the growth and development of small perturbations that potentially begin at the time of inflation. Many inflationary models give rise to a spectrum of fluctuations on scales that are larger than the cosmological horizon, and eventually these fluctuations will begin to move back into the horizon in the radiation dominated era. At this point in the Universe's development, in extreme cases, primordial black holes (PBHs) are able to form. The masses of such black holes will be very small, ranging from the Planck mass up to the horizon mass at the time of equivalence between radiation and pressureless matter. \cite{musco2005}. 

A defining characteristic of black hole formation is the event horizon, which is the boundary of the non-empty complement of the causal past of future null infinity; i.e., the region for which signals sent from the interior will never escape. For dynamical black holes, such as PBHs, we must know the global behaviour of the spacetime in order to determine the event horizon locally \cite{AshtekarKrishnan}. As an alternative, Penrose proposed the concept of {\em closed trapped surfaces without border}, which are compact spacelike surfaces such that the expansions of the future-pointing null normal vectors are negative  \cite{P2}. The {\it apparent horizon} is defined as the locus of the vanishing expansion, $\theta_{(\ell)}$ of a null geodesic congruence, $\ell_a$ emanating from trapped surfaces with spherical topology \cite{Booth2005}. The apparent horizon is quasi-local and it is intrinsically foliation-dependent.

Apparent horizons are employed in simulations of high precision waveforms of gravitational waves arising from the merger of compact-object binary systems or in stellar collapse to form black holes in numerical relativity. The observations by the LIGO collaboration of gravitational waves from black hole mergers relied upon such numerical simulations based on apparent horizons \cite{LIGO}. However, due to the foliation dependence of the apparent horizon, it is observer dependent, and this can lead to ambiguities if care is not taken to relate the differing observers' reference frames \cite{Booth2005}. For this reason it is important to determine an alternative surface that is defined invariantly, such as the {\it geometric horizon} which is a hypersurface defined by the vanishing of particular curvature invariants \cite{ADA, AD, AD2019}. 


It is of interest to determine the existence of geometric horizons for solutions describing PBH formation. The first models of PBH formation were studied in the context of spherical symmetry \cite{musco2005}, and these dynamical black hole solutions must admit geometric horizons \cite{AD}. Non-spherically symmetric PBH solutions have been considered \cite{harada2016} and it has been argued that the quasi-spherical (QS) Szekeres dust models have a more natural interpretation than the spherically symmetric solutions as a model for the formation of PBHs \cite{Harada:2015ewt}. However, if a QS Szekeres solution is to describe the formation of a PBH then shell-crossings cannot form outside of the apparent horizon, as this can be interpreted as the start of processes not described by the QS Szekeres dust solution models.

The appearance of shell-crossings arises from the choice of the metric functions \cite{hellaby02, hellaby08}. It is possible to put restrictions on the metric functions \cite{KB2012, hellaby02}, or equivalently  restrictions on the initial conditions \cite{Sussman}, in order to avoid or delay shell-crossings occurring in general \cite{gaspar2018black}. As a black hole solution, the QS Szekeres dust models require extensive fine-tuning of the black hole's mass and collapse time in order to avoid shell-crossings forming outside of the apparent horizon. If the black hole mass is within a small enough range then the time duration of collapse is ensured to be consistent with PBH formation. This suggests that a subset of the QS Szekeres dust solutions can describe the formation of PBHs in the early universe. 

The QS Szekeres dust models are known to admit an apparent horizon \cite{szekeres1975quasispherical, KB2012}. We will show that this hypersurface is, in fact, a geometric horizon \cite{ADA, AD}. To do so we will employ a frame approach to compute the appropriate Cartan invariants arising from the Cartan-Karlhede algorithm \cite{ref1, ref2a, GANG} in order to determine the existence of the geometric horizon. Previously, these models have been investigated using the orthonormal $1+3$ frame approach developed in \cite{uggla}.
This has lead to several invariant characterizations of the Szekeres dust solutions \cite{szafron1977inhomogeneous, szafron1979new, barnes1989irrotational}. The null frame approach of the Newman-Penrose (NP) formalism has been used to invariantly characterize the Szekeres solutions \cite{Wainwright77} and the Szekeres-Szafron solutions \cite{szafron1977inhomogeneous, szafron1979new}. 

From the invariant characterization of the QS Szekeres solutions,  observer based measurements of the physical properties can be described using scalar curvature invariants. Similarly, the thermodynamics of the perfect fluids of a family of the $\beta' \neq 0$ Szekeres-Szafron solutions have been considered in \cite{Coll:2018eoc} in terms of scalar invariants. While these invariants have been helpful to describe the QS Szekeres dust solutions from the perspective of inhomogeneous dust solutions, they are not well adapted to the interpretation of PBH formation.
A new set of Cartan invariants will be presented that invariantly characterize the properties of the QS Szekeres spacetimes with a physical interpretation relating to PBH formation.

The outline of the paper is as follows. In section \ref{sec:Szekeres} we review the QS Szekeres solution and discuss the spin-coefficients and curvature scalars in the NP formalism. In section \ref{sec:Cartan} the Cartan-Karlhede algorithm is applied to generate the minimal set of extended Cartan invariants. In section \ref{sec:others} we will compare the Cartan invariants with two well-known sets of scalars used to characterize the Szekeres solutions: the kinematic scalars \cite{Wainwright} and the q-scalars \cite{Sussman} to motivate the use of Cartan invariants. In section \ref{sec:properties} new extended Cartan invariants will be constructed that describe physical properties of the QS Szekeres solution and show that the apparent horizon is detected by the vanishing of a Cartan invariant. We will also construct invariants to detect shell-crossings, as their appearance outside of the apparent horizon will indicate that a given QS Szekeres solution is not a valid model for PBH formation. In section \ref{sec:examples} we will examine the zero sets of the invariants that will detect the apparent horizon and the potential appearance of shell-crossings in two examples. In section \ref{sec:discussion} we review our results and discuss future work.




\section{The quasi-spherical Szekeres dust models} \label{sec:Szekeres}

We will review the metric for the $\beta' \neq 0$ quasi-spherical (QS) Szekeres solutions with vanishing cosmological constant using the parametrization introduced by Hellaby \cite{hellaby1996null} and used in \cite{KB2012, Sussman}. We can write the metric in a simple form: 
\beq ds^2 = - dt^2 + \frac{\Ep^2 {Y'}^2}{1+2E} dz^2 + Y^2 [dx^2 + dy^2], \label{mtrc} \eeq

\noindent where $Y=Y(t,x,y,z)$ and $\Ep = \Ep(x,y,z)$ are defined as: 
\beq Y = \frac{R}{\Ep},~~ \Ep = \frac{S}{2}\left[ 1 + \left( \frac{x-P}{S}\right)^2 +\left( \frac{y-Q}{S}\right)^2\right], \label{mtrcfn} \eeq

\noindent with $R = R(t,z)$ and $S(z), P(z), Q(z)$ arbitrary functions with $S \geq 0$. Imposing the Einstein field equations with a dust source, we have the following equations:
\beq & \dot{Y}^2 = \frac{2\tilde{M}}{Y} + \tilde{E}, & \label{Yeqn}  \\
& 2 \tilde{M}' = \kappa \tilde{\rho} Y^2 Y', & \eeq
\noindent where prime and dot denotes differentiation with respect to $z$ and $t$, respectively, $ \tilde{\rho}$ is the energy density, and $$ \tilde{E} = \frac{2E}{\Ep^2},~~ \tilde{M} = \frac{M}{\Ep^3}.$$ Here, the functions $M(z)$ and $E(z)$ are called the mass and energy functions respectively \cite{Harada:2015ewt}. Expanding the first equation \eqref{Yeqn} gives a differential equation for $R$:
\beq R_{,t}^2 = 2E(z) + \frac{2M(z)}{R}. \label{dust} \eeq

\noindent The positive and negative roots determines whether the  spacetime is in the expanding or collapsing phase \cite{Booth2005}. 

We will impose the following additional conditions: 
\beq R \geq 0,~~ \text{ and } M\geq 0. \label{LTBcons} \eeq


\noindent The first is due to the interpretation of $R$ as the areal radius and hence must be positive; when $R=0$ this is either an origin, bang or crunch singularity. $M $ must be positive so that the vacuum exterior has positive Schwarzschild mass. 


In general, this solution will have no symmetry, although there are solutions which will admit rotational symmetries \cite{Nolan:2007gy, GH2017} and coordinates can be chosen so that $P$ and $Q$ are constant. With $S=1$ and $P = Q=0$, this solution reduces to the Lemaitre-Tolman-Bondi (LTB) solution, while if $R = z \tilde{S}(t)$, $E = E_0 z^2$ with $\tilde{S}$ an arbitrary function, $E_0 =$ constant, $P=Q=0$, and $S = 1$, the Robertson-Walker limit is recovered. The quasi-spherical Szekeres dust model can be regarded as a generalization of the LTB model in which the spheres of constant mass are non-concentric, with the functions $P, Q$ and $S$ determining how the center of a sphere changes its position in a space of $t = constant$ when the radius of the sphere is increased or decreased. It has been argued that these metric functions also give rise to a shell-rotation effect \cite{Buckley:2019kek}. We will assume that the metric functions are not of the form discussed in this paragraph, unless explicitly indicated.

Assuming the metric functions do not take the form of the functions discussed in the previous paragraph, we note that the sign of $E(z)$ determines the type of evolution: 
\begin{itemize}
\item If $E(z_0) < 0 $, a matter shell at $z=z_0$ expands away from the initial singularity and then recollapses to a final singularity.  
\item If $E(z_0) >0$, the shell is ever-expanding or ever-collapsing, depending on the initial conditions. 
\item If $E(z_0) = 0$, this is an intermediate case for which the shells are ever-expanding with asymptotically zero expansion, or its time-reverse.
\end{itemize}
\noindent All three evolution types can exist in different regions of the same Szekeres solution. We will consider regions where the matter is recollapsing ($E< 0$). The solution of \eqref{dust} is then \cite{KB2012}:
\beq R = - \frac{M}{2E}(1-\cos \eta),~~ \eta - \sin \eta = \frac{(-2E)^\frac32}{M} (t-t_B(z)), \label{dustsoln} \eeq
\noindent where $t_B(z)$ is an arbitrary function and $\eta(t,z)$ is a parameter.

\subsection{Spin-coefficients and curvature scalars} 

We will work with a complex null tetrad, $\{l^a, n^a, m^a {\bar m^a}\}$, such that the only non-zero inner products are $-l_a n^a=m^a {\bar m_a}=1$ and where a bar denotes a complex conjugate. In terms of the complex null tetrad the metric is then 
\beq {\bf g} =-2\ell_{(a} n_{b)} +2m_{(a} \bar{m}_{b)}, \eeq
\noindent where round parentheses denote symmetrization of indices and the tetrad is defined as
\beq  \begin{aligned} \ell_a = &\frac{1}{\sqrt{2}}\left( dt +\frac{\Ep {Y'}}{\sqrt{1+2E}} dz\right), n_a = \frac{1}{\sqrt{2}}\left( dt - \frac{\Ep {Y'}}{\sqrt{1+2E}} dz\right),  \\ 
& m_a = \frac{Y}{\sqrt{2}}( dx - i dy),~~ \bar{m}_a = \frac{Y}{\sqrt{2}}( dx + i dy). \end{aligned} \label{nllfrm} \eeq
\normalsize
\noindent We will also introduce the frame derivatives for this coframe: 
\beq \begin{aligned} & D = \frac{1}{\sqrt{2}} \left( \frac{\partial}{\partial t} - \frac{\sqrt{1+2E}}{\Ep Y'} \frac{\partial}{\partial z} \right),  \Delta  = \frac{1}{\sqrt{2}} \left( \frac{\partial}{\partial t} + \frac{\sqrt{1+2E}}{\Ep Y'} \frac{\partial}{\partial z} \right), \\
& \delta  = \frac{1}{\sqrt{2}} \left( \frac{1}{Y} \frac{\partial}{\partial x} - \frac{i}{Y} \frac{\partial}{\partial y} \right),~\bar{\delta}  = \frac{1}{\sqrt{2}} \left( \frac{1}{Y} \frac{\partial}{\partial x} + \frac{i}{Y} \frac{\partial}{\partial y} \right). \end{aligned} \eeq


The dust condition gives the following coordinate independent relations between the Ricci scalars: \beq R= 4 \Phi_{00}, \Phi_{22} = \Phi_{00} \text{ and } \Phi_{11} = \frac12 \Phi_{00},\eeq 

\noindent and the algebraically independent NP curvature scalars are:

\beq \begin{aligned}  \Phi_{00} =  \frac{\tilde{\rho}}{\kappa}  &= \frac{2\tilde{M}_{,z}}{Y^2 Y_{,z}},~ \Psi_2 = -\frac{\tilde{M}}{2Y^3}+ \frac{\kappa}{12} \tilde{\rho},~~\kappa = \frac{8 \pi G}{c^4} = 8 \pi \end{aligned}. \label{NPcurv} \eeq

\noindent That is,  the Weyl tensor is of algebraic type {\bf D}, and the Ricci tensor is of algebraic type {\bf I} relative to the alignment classification \cite{classa, classb, classc}. The divergence of the Einstein field equations gives a constraint on the energy density $\tilde{\rho}$:
\beq D\tilde{\rho} + \Delta \tilde{\rho} +  (2\epsilon +2\bar{\epsilon} + \mu + \bar{\mu} - \rho - \bar{\rho}) \tilde{\rho} = 0, \label{DivT} \eeq

%

\noindent where $\epsilon, \rho, \mu, \tilde{\kappa}$ and $\tilde{\pi}$ and their complex conjugates belong to the set of  non-zero spin-coefficients: 
\beq \begin{aligned} 
\rho &= \frac{1}{\sqrt{2}} \left( \frac{Y_{,t} \Ep - \sqrt{1+2E}}{\Ep Y} \right), \\
\mu &= -\frac{1}{\sqrt{2}} \left( \frac{Y_{,t} \Ep + \sqrt{1+2E}}{\Ep Y} \right), \\
\gamma &= -\epsilon = \frac{1}{2\sqrt{2}} \frac{Y_{,z,t}}{Y_{,z}}, \\
\tau &= \bar{\nu} = - \tilde{\kappa} = -\bar{\tilde{\pi}} = -\frac{i}{2\sqrt{2}} \left( \frac{(\Ep Y_{,z})_{,y}- i (\Ep Y_{,z})_{,x} }{\Ep Y Y_{,z}} \right). \end{aligned} \label{npend} \eeq

\section{The Cartan-Karlhede algorithm}  \label{sec:Cartan}

We will employ the Cartan-Karlhede algorithm to generate the required set of Cartan invariants for the QS Szekeres spacetime \cite{ref1, ref2a, GANG}. At each iteration, $q\geq 0$, of the algorithm, we will compute the $q$-$th$ covariant derivative of the curvature tensor and determine two discrete invariants: the number of functionally independent Cartan invariants at the $q$-th iteration, $t_q$, which are the components of the $q$-$th$ covariant derivative of the curvature tensor, and the dimension of the linear isotropy group, $dim(H_q)$, which consists of the Lorentz frame transformations that leave the curvature tensor and up to its $q$-th covariant derivative unchanged. 

Choosing a basis of functionally independent Cartan invariants, the remaining functionally independent Cartan invariants are {\em classifying functions}. Any classifying function can be expressed in terms of the functionally independent Cartan invariants, and this expression will be unchanged under coordinate transformations. Thus, if two QS Szekeres solutions have identical classifying functions, when expressed in terms of their respective functionally independent Cartan invariants, then the two QS Szekeres dust models are isometric and are related by a coordinate transformation. If any classifying function differs between the two solutions, they are distinct and there is no coordinate transformation between the two Szekeres spacetimes.

\subsection{Zeroth order Cartan invariants} 

Using the null frame \eqref{nllfrm}, the zeroth order Cartan-Karlhede algorithm can be applied readily to the Ricci and Weyl tensors. The isotropy group at zeroth order consists of spins, $m' = e^{i\theta} m$ \cite{Stewart}, and so $dim( H_0) = 1$ . In general, there are two functionally independent zeroth order Cartan invariants since the double wedge product of the exterior derivatives of $\Phi_{00}$ and $\Psi_2$ is non-zero, 
\beq d \Phi_{00} \wedge d \Psi_{2} \neq 0, \nonumber \eeq

\noindent implying that the two scalars must be functionally independent.

\subsection{ First order Cartan invariants} 

At first order, the covariant derivative of the Weyl tensor yields the following algebraically independent quantities:
\beq D\Psi_2,~\Delta \Psi_2, \delta \Psi_2, \bar{\delta}\Psi_2,~\rho, \mu, \kappa, \tau. \label{DWeyl} \eeq

\noindent While from the covariant derivative of the Ricci tensor we find additional quantities:
\beq & D \Phi_{22} + 4 \epsilon \Phi_{22}, \Delta \Phi_{22} - 4 \epsilon \Phi_{22}, \delta \Phi_{22},~~ \bar{\delta} \Phi_{22}. \label{DRicci} \eeq

\noindent The first order isotropy group is trivial, as spins  affect the form of the spin-coefficients $\kappa, \tau$, and $\epsilon$ along with any quantity differentiated by $\delta$ or its complex conjugate.  

Choosing the frame where $\epsilon$ is real-valued using an appropriate spin, this is now an invariant coframe and any frame derivative of a Cartan invariant is also a Cartan invariant. We are now able to separate the components in equations \eqref{DWeyl} and \eqref{DRicci} and work with the frame derivatives of $\Phi_{22}$ and $\Psi_2$ and the spin-coefficients directly. Assuming that the spacetime has no isometries\footnote{If a Szekeres dust-model admits a symmetry, there will only be three functionally independent invariants \cite{GH2017}, since $\epsilon \neq 0$ in these solutions. } and choosing $\epsilon$ and $\tilde{\pi}$ as the remaining two functionally independent invariants then the non-vanishing quadruple wedge product, 
\beq d \Phi_{22} \wedge d \Psi_2 \wedge d \epsilon \wedge  d \tilde{\pi} \neq 0, \nonumber \eeq

\noindent shows that the four Cartan invariants involved in the wedge product are functionally independent, and so $t_1 = 4$.

\subsection{ Second order Cartan invariants }


The Cartan-Karlhede algorithm  must continue to second order where it terminates since  $dim(H_2) = dim(H_1)=0$ and $t_2 = t_1 = 4$ \footnote{If there is a symmetry, then $dim(H_2) = dim(H_1)=0$ and $t_2 = t_1 = 3$, and so the algorithm still stops.}. The second order Cartan invariants are needed to fully characterize a given QS Szekeres dust model. 

\subsection{Kinematic quantities and q-scalars} \label{sec:others}
Choosing the timelike direction, $\sqrt{2}{\bf u} = (\bell+\bn)$,  the dust is co-moving, $u^\mu = \delta^\mu_{~0}$ and $\dot{u}^\mu = u^\nu \nabla_\nu u^\mu = 0$. The kinematic quantities along the timelike direction are: the energy density, $\tilde{\rho}$, the expansion scalar, $\Theta$, the Ricci curvature of the spatial 3-space, ${^3} \mathcal{R}$, the shear scalar, $\Sigma$, and the algebraically independent component of electric Weyl tensor, $\mathcal{W}$. These quantities  completely characterize a QS Szekeres solution \cite{Sussman}, and are related to the Cartan invariants:
\beq \tilde{\rho} &=& \frac{\kappa}{2} \Phi_{22}, \label{WE1} \\
{^3} \mathcal{R} &=& 18 \Phi_{22},\label{WE2}  \\ 
\mathcal{W} &=& 2\Psi_2, \label{WE3} \\
\Theta &=& -\frac{\sqrt{2}}{2}(2 \epsilon - \rho + \mu), \label{WE4} \\
\Sigma &=& \frac{\sqrt{2}}{6} (4 \epsilon + \rho - \mu) \label{WE5} . \eeq

Expansion-normalized variables can be constructed from these scalars that give dimensionless evolution equations dictating the dynamics of the Szekeres spacetime as a set of scalar evolution equations, a `Hamiltonian' constraint and spacelike constraints \cite{Wainwright, Ellis}. Since $\bell+\bn = -\sqrt{2}\partial_t$, the equation arising from the divergence of the Einstein field equations \eqref{DivT} can be rewritten in terms of these quantities: 
\beq \tilde{\rho}_{,t} +  \Theta \tilde{\rho} = 0, \eeq
\noindent which agrees with the first scalar evolution equation in equation (17) of \cite{Sussman}. The Raychaudhuri equation and the remaining evolution equations are then:
\beq \Theta_{,t} &=& -\frac{\Theta^2}{3} - \frac{\kappa}{2} \tilde{\rho}-6 \Sigma^2, \label{Rayc} \\
\Sigma_{,t} &=& -\frac23 \Theta \Sigma - \Sigma^2 + \mathcal{W}, \label{Sevo}\\
\mathcal{W}_{,t} &=& - \Theta \mathcal{W} - \frac{\kappa}{2} \tilde{\rho} \Sigma + 3 \Sigma \mathcal{W}. \label{Wevo} \eeq 

\noindent The ``Hamiltonian'' constraint and spacelike constraints are, respectively\footnote{ Since $u_a = dt$, the projection operator $h_{ab} = g_{ab} + u_a u_b =  g_{ab} + 2(\ell_a + n_a) (\ell_b + n_b)$ was used to compute the Ricci scalar, ${^3} \mathcal{R}$, of the hypersurfaces $t=const$. To recover the form in \cite{Sussman} we notice that $\dot{Y} = 0$ and so $\tilde{M} = \frac12 \tilde{K} Y$.}:
\beq & \frac{\Theta^2}{9} = \frac{\kappa \tilde{\rho}}{3} - \frac{{^3} \mathcal{R}}{6} + \Sigma^2, & \\
& \tilde{\nabla}_b \sigma^b_a - \frac23 h^b_a \Theta_{,b} = 0,~~ \tilde{\nabla}_b W^b_a - \frac{\kappa}{3} h^b_a \tilde{\rho}_{,b} = 0. & \eeq
\noindent Relative to the invariant frame determined by the Cartan-Karlhede algorithm, the above equations can be expressed in terms of zeroth and first order Cartan invariants \cite{GANG}.

There is another set of coordinate independent scalar variables   which are quasi-local and are defined in terms of appropriate integral distributions of the local kinematic variables, $\{ \tilde{\rho}, {^3}\mathcal{R}, \mathcal{W}, \Theta, \Sigma\}$, giving the set of q-scalars: $$\{ \rho_q,\mathcal{H}_q, \mathcal{K}_q, \Sigma_q\}.$$ The q-scalars can be interpreted as weighted averages of the local scalars when treated as functionals. The local kinematic scalars can then be treated as fluctuations of the q-scalars: $$\{ \Delta^{(\rho)}, \Delta^{(\mathcal{H})}, \Delta^{(\mathcal{K})}, \Delta^{(\Sigma)}  \}.$$ 

Due to the relationship between the original kinematic scalars and the q-scalars, the evolution equations for the q-scalars can be rewritten in terms of the original quantities $\tilde{\rho},~ \Theta,~ \Sigma,~ \mathcal{W}$ and ${^3} \mathcal{R}$ \eqref{WE1}-\eqref{WE5}. For example, considering the evolution equations \cite{Sussman}: 
\beq \dot{\rho}_q &=& -3\rho_q \mathcal{H}_q, \\
\dot{ \mathcal{H}_q} &=& -  \mathcal{H}_q^2 - \frac{\kappa}{6}\rho_q, \\
\dot{\Delta}^{(\rho)} &=& - 3(1+ \Delta^{(\rho)})  \mathcal{H}_q \Delta^{(\mathcal{H})}, \\
\dot{\Delta}^{(\mathcal{H})} &=& -(1+3\Delta^{(\mathcal{H})})  \mathcal{H}_q \Delta^{(\mathcal{H})} + \frac{\kappa \rho_q}{6 \mathcal{H}_q} (\Delta^{(\mathcal{H})} - \Delta^{(\rho)}), \eeq
\noindent then the first two equations, expressed in terms of $\tilde{\rho}$ and $\Theta$ agree with \eqref{DivT}-\eqref{Wevo} and the Hamiltonian and spatial  constraints become:
\beq & \mathcal{H}_q^2 = \frac{\kappa}{3} \rho_q - \mathcal{K}_q, & \label{SussmanHq}\\
& 2 \Delta^{(\mathcal{H})} = \Omega_q \Delta^{(\rho)} + (1 - \Omega_q) \Delta^{(\mathcal{K})}, \label{SussmanDH}& \eeq

\noindent where $\Omega_q$ is a q-scalar analogue of the FLRW Omega factor $\Omega = \Omega_q (1 + \Delta^{(\Omega)})$, with its corresponding fluctuation: 
\beq & \Omega_q = \frac{\kappa \rho_q}{3 \mathcal{H}_q^2},~~ \Omega_q -1 = \frac{\mathcal{K}_q}{\mathcal{H}_q^2},& \label{SussmanOq} \\ 
& \Delta^{(\Omega)} = \Delta^{(\rho)} - 2 \Delta^{(\mathcal{H})} = (1 -\Omega_q) (\Delta^{(\rho)} - \Delta^{(\mathcal{K})}). \label{SussmanDO} & \eeq

We note that the q-scalars $\rho_q,\mathcal{H}_q, \mathcal{K}_q$ and $\Sigma_q$, and their fluctuations as determined in \cite{Sussman}, can be expressed in terms of the Cartan invariants through the expressions \eqref{WE1}-\eqref{WE5} and the identities given in Appendix B of \cite{Sussman}. For example, 
\beq &\rho_q = \frac{6 \Psi_2 + \kappa \tilde{\rho}}{\kappa},~~\Delta^{(\rho)} = \frac{6\Psi_2/(-\kappa \tilde{\rho})}{1-6\Psi_2/(-\kappa \tilde{\rho})} \label{qsc1}& \\
& \mathcal{H}_q = 3 \Theta + \Sigma,~~ \Delta^{(\mathcal{H})} = -\frac{\Sigma/(3\Theta)}{1+\Sigma/(3\Theta)}. \label{qsc2} & \eeq

In a similar manner, the q-scalars $ \mathcal{K}_q$ and $\Sigma_q$, and their fluctuations which are derived by applying the constraints the equations \eqref{SussmanHq} - \eqref{SussmanDH} and \eqref{SussmanOq} - \eqref{SussmanDO} to \eqref{qsc1} and \eqref{qsc2}, respectively,  will also be extended Cartan invariants.

\section{Invariant Characterization of Physical Properties} \label{sec:properties}

The QS Szekeres solutions can describe an inhomogeneous cosmological model \cite{Wainwright77}, a wormhole solution \cite{hellaby02} or the formation of a primordial black hole \cite{Harada:2015ewt,gaspar2018black}. The interpretation of a QS Szekeres solution is dependent on the behaviour of particular properties associated with the geometry, which can be considered as physical characteristics. For example, the appearance of shell-crossings before the apparent horizon forms or outside of the apparent horizon are geometric properties that immediately exclude a QS Szekeres solution as a model for PBH formation.  

Even within the class of QS Szekeres solutions which describe the formation of PBHs, the behaviour of these physical properties will be important. For example, the conditions for the formation of  future and past apparent horizons in  Szekeres-Szafron spacetimes  depend on the expansion or contraction of spacetime, the location of shell-crossings and the relative movement between matter shells \cite{polavskova2018}. When the apparent horizon exists, we will show that it is a geometric horizon. While we are not primarily concerned with the properties of apparent horizons, we believe that the invariant characterization of these properties will give insight into the physical interpretation of the geometric horizon.


\subsection{Detection of the horizon}

The QS Szekeres dust models admit an apparent horizon, defined by the surface $R=2M$, which corresponds to the vanishing expansion of the future-pointing null vector normal to this surface \cite{KB2012}. Due to the lack of a timelike Killing vector or spherical symmetry, there are no previously known scalar polynomial curvature invariants (SPIs) that will, in general, detect the apparent horizon \cite{PageShoom2015, faraoni2017foliation}.

To detect the apparent horizon, we will consider the covariant derivative of the Weyl tensor. The components of $C_{abcd;e}$ may be expressed in terms of $\Psi_2$, $\Phi_{11}$, $\Delta \Phi_{11}$ and the spin-coefficients \eqref{npend}. In the chosen invariant coframe, the form of $C_{abcd;e}$ does not conform with the known algebraic types from the alignment classification \cite{classa, classb, classc} \footnote{Since the discriminant SPIs built from the Weyl and Ricci tensors, along with the covariant derivatives of these tensors do not vanish anywhere, these tensors cannot be of alignment type {\bf II} or more special.}. Using the  algebraic and differential Bianchi identities, the non-zero components of $C_{abcd;e}$ are:
\beq C_{1214;3} = C_{1434;3} = C_{1213;4} = C_{1334;4} = 3 \rho \Psi_2 \label{SzkBW10} \eeq
\noindent and $2C_{1423;1} = C_{1212;1} = C_{3434;1}$ where
\beq C_{3434;1} =  \frac{-2\Delta  \Phi_{11} - 32 \epsilon \Phi_{11} -4 \mu \Phi_{11}+ \rho( 18 \Psi_2+ 4  \Phi_{11})  }{3}. \label{SzkBW01} \eeq 

To show that $R=2M$ is a geometric horizon, we note that the extended Cartan invariant $\rho$, defined in equation \eqref{npend}, will vanish on the surface $R=2M$ \cite{ADA, AD}. This surface is a dynamical geometric horizon since the extended invariant, $\mu$, which also appears in the covariant derivative of the curvature tensor, is negative within the surface $R=2M$  \cite{AD2019}. The spin-coeficients $\rho$ and $\mu$ correspond to the expansion of the ingoing and outgoing null directions; i.e., $\rho = \theta_{(\ell)}$ and $\mu = \theta_{(n)}$.

We have only considered the contracting phase by choosing the negative root of \eqref{dust}; however, in the expanding phase, where \eqref{dustsoln} has a positive root, $\mu$ vanishes on $R=2M$ and $\rho$ is negative within this invariant surface, implying that the expanding phase admits a dynamical geometric horizon. 

We note that the geometric horizon coincides with the apparent horizon; however, the geometric interpretation of these two surfaces differs. In the case of a geometric horizon the preferred null directions $\ell_a$ and $n_a$ in \eqref{nllfrm} are not geodesic but lie in the timelike plane spanned by the null normals to the hypersurface $R=2M$ \cite{AD}. In general, geometric horizons are not necessarily apparent horizons or any other horizon based on trapped surfaces \cite{AD2019}. 

\subsection{Areal radius}

The areal radius can be shown to be an invariant function expressed in terms of the Cartan invariants. By first combining the algebraically independent Ricci scalar and Weyl scalar, we can write
\beq C_0 = \Psi_2 - \frac{\kappa}{12} \tilde{\rho} = -\frac{\tilde{M}}{2 Y^3} = -\frac{M}{2R^3}. \label{C1eqn} \eeq
Then using $\rho$ and $\mu$ in \eqref{npend}, we have two simpler invariants
\beq \begin{aligned} C_1 &= -\frac{\sqrt{2} (\rho+\mu)}{2} = \frac{\sqrt{1+2E}}{R}, \\ C_2 &= \frac{\sqrt{2} (\rho-\mu)}{2} = \frac{Y_{,t} \Ep }{\Ep Y} = \frac{ R_{,t}}{ R}. \\ \end{aligned} \label{C1C2eqn} \eeq

\noindent The differential equation \eqref{dust} allows us to combine the Cartan invariants:
\beq -4 C_0 + C_1^2 -C_2^2 = \frac{2M}{R^3}+\frac{1+2E}{R^2} - \frac{R_{,t}^2}{R^2} = \frac{R_{,t}^2}{R^2} + \frac{1}{R^2}  - \frac{R_{,t}^2}{R^2} = \frac{1}{R^2}. \nonumber \eeq

\noindent Outside of the singularities that occur when $R=0$ (a bang, a crunch or potentially the origin), $R>0$ and so the areal radius can be isolated by taking the square root. It is possible to choose initial conditions so that the origin is regular \cite{hellaby02}.


\subsection{The mass and energy functions}

The mass function, $M$, and energy function, $E$, are Cartan invariants as well since
\beq M = -2 C_0 R^3, \text{ and } E = \frac{C_1^2 R^2 - 1}{2}. \eeq  

\subsection{Expansion or contraction of spacetime}

The expansion or contraction of spacetime is an invariant quantity determined by the difference of $\rho$ and $\mu$:
\beq C_2 &= \frac{\sqrt{2} (\rho-\mu)}{2} = \frac{Y_{,t} \Ep }{\Ep Y} = \frac{ R_{,t}}{ R}. \eeq

\subsection{Spatial extrema and rate of change of the areal radius}

As the invariant frame derivative operators can be added together to give a new operator proportional to the coordinate derivative,
\beq D_z = \frac{D + \Delta }{\sqrt{2}} = \frac{\sqrt{1+2E}}{\Ep Y_{,z}} \partial_z, \label{zderiv} \eeq
we can apply this to the inverse square of the areal radius to produce a new extended Cartan invariant:
\beq D_z R^{-2} =- \frac{\sqrt{1+2E} R_{,z}}{\Ep Y_{,z} R^3}. \eeq
\noindent This invariant will detect the spatial extrema, $R_{,z} = 0$, when the numerator vanishes and indicate distinct regions where $R_{,t}$ changes sign.

We can determine the sign of $R_{,t}$ in a region using the fact that $\tilde{\rho} \geq 0$ and
\beq {\tilde{\rho}} = \frac{2 \Ep^3 \tilde{M}_{,z}}{R^2 \Ep Y_{,z}} =  \frac{2}{R^2} \left(\frac{M_{,z} - 3\frac{M \Ep_{,z}}{\Ep} }{R_{,z}-\frac{R \Ep_{,z}}{\Ep}} \right), \label{BigED} \eeq
\noindent  in regions where $R_{,z} \neq 0 $ and $0 < \tilde{\rho} < \infty$ we can compare the sign of the following invariant,
\beq \frac{D_z R}{\tilde{\rho}} = \frac{Y^2 \sqrt{1+2E} R_{,z}}{2 \Ep \tilde{M}_{,z}} = \frac{R^2 \sqrt{1+2E} R_{,z}}{2 \Ep^3 \tilde{M}_{,z} }.  \label{inv:DzRoED} \eeq

\noindent Since $\Ep Y_{,z}$ and $\Ep^3 \tilde{M}_{,z}$ must both be positive or negative in the same region, the change in the sign of $R_{,z}$ can be determined. We note that the Szekeres metric is covariant under the transformation $r = g(\tilde{r})$. If $R_{,z} < 0$ then new local coordinates can be chosen so that $R_{,z} >0$ \cite{hellaby02}. However, if in a region $R_{,z}$ changes sign, this is a coordinate independent property. 

\subsection{Spatial extrema and rate of change of the mass function}

Consider the invariant derivative of the mass function with respect to the derivative operator \eqref{zderiv}:
\beq D_z M = \frac{\sqrt{1+2E} M_{,z}}{\Ep Y_{,z}}. \label{inv:mzderiv} \eeq

\noindent This will detect the spatial extrema of the mass function. In regions where $M_{,z} \neq 0 $ and $0 < \tilde{\rho} < \infty$ the sign of $M_{,z}$ can be determined by comparing the sign of another invariant,
\beq C_3 = \frac{D_z M}{\tilde{\rho}} = \frac{Y^2 \sqrt{1+2E} M_{,z}}{2 \Ep \tilde{M}_{,z}} = \frac{R^2 \sqrt{1+2E} M_{,z}}{2 \Ep^3 \tilde{M}_{,z} }. \label{inv:c3} \eeq

\noindent Due to equation \eqref{BigED}, the expressions $\Ep Y_{,z}$ and $\Ep^3 \tilde{M}_{,z}$ must both be either positive or negative in the same region and this determines the sign of $M_{,z}$.

\subsection{Shell-crossings}

When two matter shells with different $z$-values move towards each other and intersect, a shell-crossing occurs and this leads to a weak curvature singularity since $\tilde{\rho}$ diverges at the location of a shell-crossing \cite{hellaby02}. To determine when this occurs, we can define the distance between shells locally by \beq \sqrt{g_{zz}} = \frac{\Ep Y_{,z}}{\sqrt{1+2E}} = \frac{R_{,z}-\frac{R \Ep_{,z}}{\Ep}}{\sqrt{1+2E}} \label{Shellmetric}, \eeq

\noindent Thus, a shell-crossing occurs when the numerator is zero and the energy density $\tilde{\rho}$ diverges. As this is a curvature singularity, it will be reflected in the curvature invariants. We would like to find invariants that vanish and do not diverge when a shell-crossing occurs, we will consider the inverse of the energy density:
\beq \frac{1}{\tilde{\rho}} = \frac{Y^2 Y_{,z}}{2 \tilde{M}_{,z}} =  \frac{R^2}{2} \left(\frac{R_{,z}-\frac{R \Ep_{,z}}{\Ep}}{M_{,z} - 3\frac{M \Ep_{,z}}{\Ep} } \right). \label{inv:1oED} \eeq

In principle, the denominator of  $\tilde{\rho}$ will vanish in two cases: when a shell-crossing occurs or when the numerator vanishes as well. If the numerator and denominator vanishes, this is called {\it a neck or belly} and this will be discussed in subsection \ref{subsec:neck}. In order to distinguish between a shell-crossing and a neck, one must compute an additional extended Cartan invariant with the derivative operator \eqref{zderiv}:
\beq C_4 = \left[ D_z M \right]^{-1} = \frac{\Ep Y_{,z}}{\sqrt{1+2E} M_{,z} }. \label{scdetect1} \eeq
\noindent 

If a shell-crossing exists, then this surface can be determined by the vanishing of two invariants: \beq \tilde{\rho}^{-1} = 0 \text{ and } C_4 = 0.\eeq We note that these hypersurfaces may not entirely intersect with the $r=constant$ 2-spheres for a given value of $t$ due to the $x$ and $y$ dependence in the numerators of $\tilde{\rho}^{-1}$ and $C_4$. 



\subsection{Movement of the matter shells}

To determine the relative motion of matter shells, we can differentiate the local distance $\sqrt{g_{zz}}$ by $t$ to give:
\beq ( \sqrt{g_{zz}})_{,t} = \frac{(\Ep Y_{,z})_{,t}}{\sqrt{1+2E}}, \nonumber \eeq

\noindent from which it can be determined whether matter shells are, respectively, moving away or moving together for $$( \sqrt{g_{zz}})_{,t} > 0 \text{ or } ( \sqrt{g_{zz}})_{,t}<0.$$ 

In order to construct Cartan invariants that invariantly describe this behaviour, we will combine $\Sigma$ and $\Theta$ in \eqref{WE4} and \eqref{WE5} to construct another extended Cartan invariant $\epsilon$:
\beq  \epsilon = -\frac{1}{2\sqrt{2}} \ln (Y_{,z})_{,t}. \nonumber \eeq

\noindent Multiplying $\epsilon$ by $C_4$ yields the required invariant: 
\beq C_5 = \epsilon C_4 = \frac{(\Ep Y_{,z})_{,t}}{\sqrt{1+2E} M_{,z} }. \nonumber \eeq

\noindent Since the sign of $M_{,z}$ can be determined invariantly, the sign of $(\Ep Y_{,z})_{,t}$ is given by the sign of the Cartan invariant $C_5$.
 
\subsection{Necks and Bellies: Regular Maxima and Minima} \label{subsec:neck}

The spatial slices of a QS Szekeres solution can have spatial extrema in $R$: either a maximum  areal radius for closed spatial sections or a minimum areal radius for wormholes \cite{krasinski2004}. It is possible that the QS Szekeres solution will become degenerate or singular at points. For example, in the LTB limit, the equality $1+2E = R_{,z}^2 = 0 $ can occur when $R_{,z} = 0$ at $z=z_m$. To ensure the metric components are finite at $z=z_m$ we require that $\Ep Y_{,z} = 0$, which then implies that $\Ep^3 \tilde{M}_{,z} = 0$ to keep the energy density finite. If these conditions hold at $z=z_m$, then the following equations must hold:
\beq M_{,z} - \frac{3M \Ep_{,z}}{\Ep}=0,~~R_{,z} - \frac{R \Ep_{,z}}{\Ep}=0. \nonumber \eeq
\noindent The surface defined by $r=r_m$ is either a regular minima (a neck) or a regular maxima (a belly). 


To determine when either a neck or belly occurs, we must determine when 
\beq \tilde{\rho} = 0 \text{ and } C_4 = 0. \label{neckdetect} \eeq
\noindent The change of sign of $R_{,z}$ on either side of $z=z_m$ determines if the regular extrema is a neck or belly.





\section{Examples} \label{sec:examples}

\subsection{A model for galactic black hole formation}

As a simple example of the collapsing QS Szekeres dust models with no cosmological constant, we will consider the example given in \cite{bolejko}. This is a special case of the QS Szekeres dust models which can be seen as a generalization of the LTB models describing the formation of galactic-sized black holes without any shell-crossings. We will choose coordinates where $\tilde{z} = M(z)$, effectively setting $z = M$ in the new coordinate system: 
\beq t_B(M) = -bM^2 + t_{B0},~~ t_C(M) = aM^3 + T_0 + t_{B0} , \eeq

\noindent where $t_B(M)$ is the big bang time, $t_C(M)$ is the crunch time, and $a,b,t_{B0}$ and $T_0$ are arbitrary constants. In particular, $t_{B0}$ is the time-coordinate of the central point of the big bang and $T_0$ is the time between the big bang and the big crunch measured along the central line $M=0$. Since $\eta = 2\pi$ at $t=t_C$, this gives a simple form for $E$:
\beq 2 E(M) = - \frac{(\kappa M)^{\frac23}}{4^{\frac23}(aM^3 + bM^2 + T_0)^{\frac23}}.  \eeq

\noindent To ensure that the mass density is not negative or infinite at any point in space, we will employ the following parameter values and functions:
\beq \begin{aligned} & a=0.1, b=5000, T_0 = 12.5, t_{B0} = 0, S = M^{0.29}, P = 0.5 M^{0.29},~~Q=0. & \end{aligned} \eeq

\begin{figure}[h!] 
  \centering
\begin{subfigure}{0.5\textwidth}
    \includegraphics[scale =0.4]{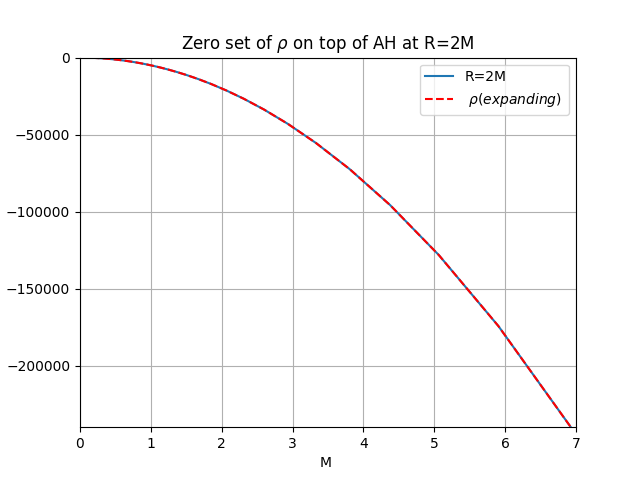}
\end{subfigure} 
\begin{subfigure}{0.4\textwidth}
    \includegraphics[scale =0.4]{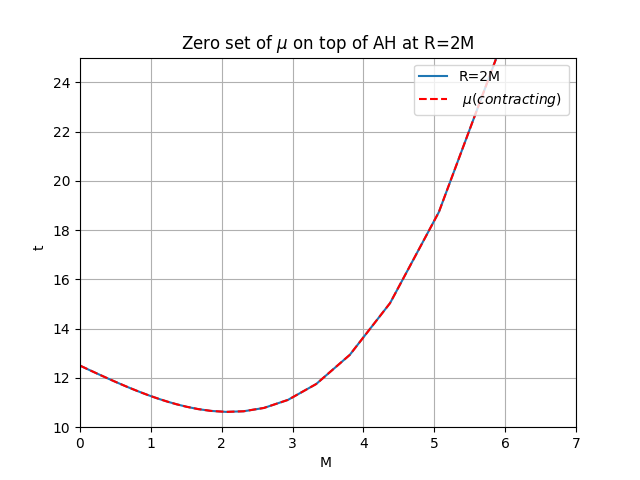}
    \end{subfigure}  
      \caption{Example 1: The zero-set of the Cartan invariant (blue) $\rho\mu$ for the expanding phase $R_{,t} >0$ (left) and for the contracting phase $R_{,t}<0$ (right) in the $(M,t)$ plane compared with the equation for the apparent horizon $R=2M$ (red). } \label{fig:example1}
\end{figure}

The metric functions have been chosen so that shell-crossings never occur. This is reflected in the Cartan invariants $C_3$ and $C_4$ which are always non-zero. The apparent horizon can be determined by plotting the values for $t$ and $z$ where the extended Cartan invariant $\rho$ (or $\mu$) vanishes as displayed in figure \ref{fig:example1}.

\subsection{Formation of a primordial black hole}

We will now consider model D in \cite{Harada:2015ewt} as an example of the formation of a primordial black hole with shell-crossing singularities. In this model, the shell-crossings will be contained within the apparent horizon. The QS Szekeres solution is generated from a reference LTB solution by adding an axisymmetric dipole to deviate from spherically symmetry. The functions in \eqref{dustsoln} are:
\beq \begin{aligned} E(z) &= \begin{cases} - \frac12 \left(\frac{z}{r_c}\right)^2 \left(1+\left(\frac{z}{r_w}\right)^{n_1}- 2\left(\frac{z}{r_w}\right)^{n_2}\right)^4 & 0< z < r_w \\ 0 & z \geq r_w  \end{cases},\\ 
M(z) &= \frac12 z^3, \end{aligned} \eeq

\noindent where $r_c$ and $r_w$ are positive constants, $n_1>1$ and $n_2>2$ are positive integers and for consistency $r_c/r_w > \sqrt{f_{max}}$ must be satisfied where $f(x) = x^2(1+x^{n_1} -2x^{n_2})^4$ for $0 < x <1$. We will choose the constants: 
$$ n_1 = 8,~n_2 = 10,~r_c = 10,~\text{and } r_w =1.$$
\noindent To deviate from spherical symmetry we will choose $P(z) = Q(z) = 0$ and $S(z) = \sqrt{2} z$. 

\begin{figure}[h!] 
  \centering
\begin{subfigure}{0.5\textwidth}
    \includegraphics[scale =0.4]{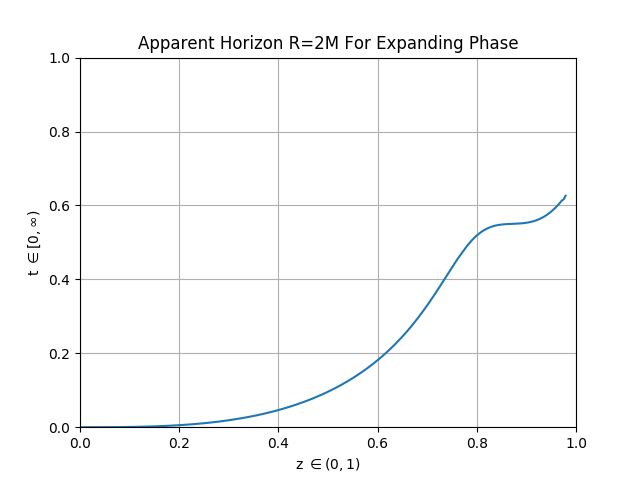}
\end{subfigure} 
\begin{subfigure}{0.4\textwidth}
    \includegraphics[scale =0.4]{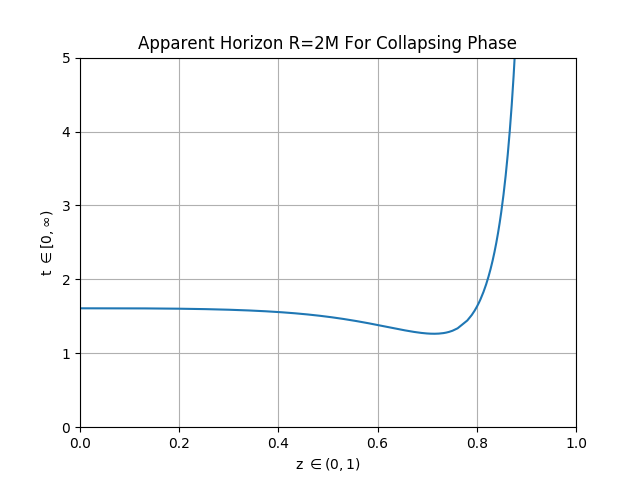}
    \end{subfigure}  
      \caption{Example 2: The zero-set of the Cartan invariant $\rho\mu$ for the expanding phase $R_{,t} >0$ (left) and for the contracting phase $R_{,t}<0$ (right). } \label{fig:example2a}
\end{figure}

The behaviour of the apparent horizon of the full QS Szekeres solution is displayed in figure \ref{fig:example2a} by graphing the zero-sets of $\mu$ or $\rho$ for the expanding and collapsing phases, respectively.

In the expanding phase, the pair of Cartan invariants $C_3$ and $C_4$ are always non-zero, implying that there are no shell-crossing singularities in the expanding phase. In the collapsing phase the Cartan invariants $C_3$ and $C_4$ admit a non-trivial zero-set. To graph the occurrence of shell-crossings of the QS Szekeres solution, we must consider the zero sets of $C_3$ and $C_4$ and graph the resulting surfaces in three-dimensions for a chosen set of $t=constant$ slices. 

We note that any shell-crossing that occurs in the reference LTB solution will form at a later time in the QS Szekeres solution, once the dipole has been reintroduced  \cite{Harada:2015ewt}.  While in practice one will consider the full QS Szekeres solution instead of the reference LTB solution, the latter provides the advantage that the shell-crossings can be graphed in 2D which clearly shows the formation of shell-crossings after the apparent horizon forms. 

In the collapsing phase of the LTB seed, we see that a shell-crossing forms at a particular value of $z$ and $t$ and persists for the remainder of the solution. However, the shell-crossing appears within the region bounded by the surface $R=2M$ and will not interact with the exterior region.

\begin{figure}[h!] 
  \centering
    \includegraphics[scale =0.5]{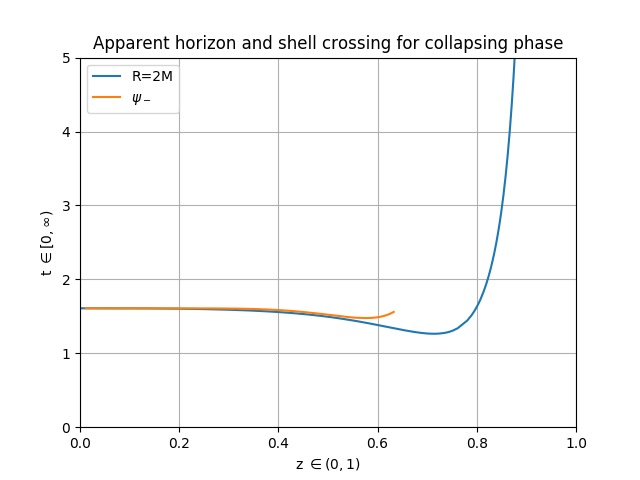}
      \caption{Example 2: The zero set of $\Psi$, the numerator of $C_4$, in orange for the LTB reference solution indicating a shell-crossing will occur. The geometric horizon is displayed in blue for comparison.} \label{fig:example2b}
\end{figure}

\begin{rem} 
The determination of the shell-crossing singularities for the QS Szekeres solution and the reference LTB solution are computationally comparable. We have chosen to work with the LTB solution purely for the purposes of displaying the graph in figure \ref{fig:example2b}. 
\end{rem}

\section{Discussion} \label{sec:discussion}

We have considered the role of QS Szekeres dust solutions as potential black hole solutions and introduced an invariant characterization of the apparent horizon. While the family of QS Szekeres solutions always admit an apparent horizon, only a subset of these solutions permit a physical interpretation as black hole solutions due to particular features that can occur during the evolution of a QS Szekeres solution, such as the appearance of shell-crossing singularities outside of the apparent horizon, or the duration until collapse of a QS Szekeres black hole being lesser or greater than the anticipated values. Due to the short time of collapse for QS Szekeres solutions it is argued that they are well-suited to describing the formation of primordial black holes in the early universe. 

We have shown that relative to an appropriate coframe coinciding with the preferred timelike direction ${\bf u}$ \cite{Wainwright77}, the kinematic scalars and the q-scalars along with their respective  evolution equations (both of which fully characterize the QS Szekeres dust solutions) can be expressed in terms of Cartan invariants. While the kinematic scalars and q-scalars describe the evolution of QS Szekeres solutions irrespective of their interpretation, they are unable to identify geometric characteristics (such as an invariant characterization of the apparent horizon) that would be suitable for the interpretation of a black hole solution. 

To address this, a new set of curvature invariants has been introduced that are adapted to the interpretation of QS Szekeres PBH solutions. In addition to showing that the apparent horizon is detected by the vanishing of a Cartan invariant, implying that it is a geometric horizon \cite{ADA, AD, AD2019},  we have also introduced invariants to characterize the expansion or contraction of spacetime itself, the spatial rate of change and extrema of the areal radius, the spatial rate of change and extrema of the mass function, the relative movement of matter shells, the existence of shell-crossings and regular spatial extrema in a QS Szekeres solution. We note that this new set of invariants can describe the evolution of any QS Szekeres dust models and has a physical interpretation. These physical properties can distinguish whether a given QS Szekeres solution is a valid model for galaxy formation, a wormhole or the formation of a PBH. 

The geometric horizon could be helpful in addressing the possibility of global visibility in these spacetimes, which occurs when a light ray emanates from the singularity before the event horizon forms \cite{Harada:2015ewt}. In asymptotically  flat spacetimes, we can define globally naked singularities in terms of future null infinities. In the cosmological setting this is not well defined, and so in order to investigate the global visibility, by determining the event horizon, null radial geodesics which emanate from the singularity must be tracked. This is a difficult problem because null geodesics cannot be kept radial and the null geodesic equations cannot be integrated analytically in general.

For both the general spherically symmetric metric and the QS Szekeres dust solutions, the discriminant SPIs built from the Weyl and Ricci tensors, along with their covariant derivatives, are non-zero on the apparent horizon (although combinations of them can vanish). Relative to the coframe chosen from the Cartan-Karlhede algorithm there appears to be some regular structure in the covariant derivatives of the Weyl tensor. This suggests that for dynamical black hole solutions the covariant derivative of the Riemann tensor will be algebraically special on a geometric horizon but it will not necessarily be readily classified using the alignment classification, and hence may not necessarily be of type {\bf II}. 

At this stage there is no procedure to generate a SPI which is globally defined and vanishes on the geometric horizon. However, it is possible to locally solve for the relevant Cartan invariants in terms of SPIs and this is reviewed in the appendix. Fortunately, there is a Cartan invariant, $\rho$ (or $\mu$), that will indicate the existence of the geometric horizon  and in the case of the spherically symmetric metric a SPI  will detect it \cite{AD}. Therefore, $\rho = 0$ (or $\mu=0$) provides a putative characterization for the geometric horizon. 
The vanishing of $\rho$ (or $\mu$), relative to the invariant coframe, is an integral part of the definition of a geometric horizon. This condition will be examined for more general Szekeres dust models, such as the QS Szekeres solutions with non-zero cosmological constant, and warrants further investigation for less idealized solutions in GR.  It is also of interest to explore the relationship between the appearance of $\rho$ at first order and spacetimes admitting a tensor of alignment type {\bf D}.

%


\section*{Acknowledgements} 
We would like to thank Ismael Delgado Gaspar and  Daniele Gregoris for useful discussions at the beginning of this project. The work was supported by NSERC of Canada (A.C.), and through the Research Council of Norway, Toppforsk grant no. 250367: Pseudo-Riemannian Geometry and Polynomial Curvature Invariants: Classification, Characterisation and Applications (D.M.). 

\subsection*{Appendix: Frame independent curvature invariants}
As in the case of the spherically symmetric metrics, the components in \eqref{SzkBW10} vanish on the apparent horizon $R = 2M$, while the components in \eqref{SzkBW01} do not. This relationship is reflected in the vanishing of the Cartan invariant  $\rho$ relative to the invariant coframe chosen by the Cartan-Karlhede algorithm. Taking the zeroth order and first order SPIs: 

\beq I_1 = C_{abcd} C^{abcd} = \Psi_2,~~R = R^a_{~a} = 8 \Phi_{11}, \label{SPIa}\eeq 

\noindent along with the quadratic first order SPIs: 
\beq \begin{aligned}& I_3 = C_{abcd;e} C^{abcd;e},~~I_{3a} = C_{abcd;e} C^{ebcd;a}, I_5 = I_{1;a}I_1^{~;a},  \\ 
&J_1 = R_{ab;c} R^{ab;c},~~ J_2 = R_{ab;c} R^{ac;b},~~J_3 = R_{;a}R^{;a}, \end{aligned} \label{SPIb} \eeq 

\noindent we can produce the following algebraically independent SPIs:

\beq & (\mu - \rho)(\mu - \rho + 8 \epsilon), & \nonumber \\
& \epsilon( \mu - \rho - \epsilon), & \nonumber\\
& \rho \mu  - 2 |\tau|^2, & \nonumber \\
& \mu \Delta \ln (\Phi_{11}) + 4 \rho \Delta \ln (\Phi_{11}) +  8 \rho \mu + 16 \rho \epsilon - 8 \rho^2 - 9 \rho \mu \frac{\Psi_2 }{ \Phi_{11}}, & \nonumber\\
& 2^2 5 \rho \Delta \ln (\Phi_{11}) - 2^5 \epsilon \Delta \ln (\Phi_{11}) + 2^5 \rho \mu + 2^6 \rho \epsilon + 2^5 \rho^2 - 6^2 \rho \mu \frac{\Psi_2}{\Phi_{11}} + 3^2 \rho \mu \frac{\Psi_2^2}{\Phi_{11}^2}, & \nonumber \\ 
& 2^7 (\Delta \ln (\Phi_{11})^2 - 2^8 \rho \Delta \ln (\Phi_{11}) + 2^8 \mu \Delta \ln (\Phi_{11}) + 2^9 \epsilon \Delta \ln (\Phi_{11}) + 6^2 2^3 |\tau|^2 \frac{\Psi_2^2}{\Phi_{11}^2}. & \nonumber \eeq

\noindent The six SPIs in \eqref{SPIa} and \eqref{SPIb} are polynomials in terms of six Cartan invariants: $$\Delta \ln (\Phi_{11}), \rho, \mu, \epsilon, |\tau|^2,~~and~~ \frac{\Psi_2}{\Phi_{11}}.$$ \noindent Locally, it is possible to express $\rho$ (or $\mu$) as a function of these SPIs in order to detect the horizon when the Jacobian of these polynomials in terms of the six Cartan invariants is non-zero. However, this will introduce additional regions where the SPIs will vanish, giving rise to the possibility of the incorrect detection of the apparent horizon.

\bibliographystyle{unsrt-phys}
\bibliography{SzekeresRefs}

\end{document}